\documentclass[aps,10pt,prb,twocolumn,showpacs,amsmath,amssymb]{revtex4}
\usepackage{graphicx}
\usepackage{color}
\newcommand{\ud}{\mathrm{d}}

\begin{document}

\title{Thermoelectricity in a junction between interacting cold atomic Fermi gases}

\author{Tibor Sekera}
\email[Email address: ] {tibor.sekera@unibas.ch}
\author{Christoph Bruder}
\affiliation{Department of Physics, University of Basel,
Klingelbergstrasse 82, CH-4056 Basel, Switzerland}
\author{Wolfgang Belzig}
\affiliation{Fachbereich Physik, Universit\"at Konstanz, 
D-78457 Konstanz, Germany}
\date{\today}
 
\begin{abstract}
  A gas of interacting ultracold fermions can be tuned into a strongly
  interacting regime using a Feshbach resonance.  Here we
  theoretically study quasiparticle transport in a system of two
  reservoirs of interacting ultracold fermions on the BCS side of the
  BCS-BEC crossover coupled weakly via a tunnel junction. Using the
  generalized BCS theory we calculate the time evolution of the system
  that is assumed to be initially prepared in a non-equilibrium state
  characterized by a particle number imbalance or a temperature
  imbalance. A number of characteristic features like sharp peaks in
  quasiparticle currents, or transitions between the normal and
  superconducting states are found. We discuss signatures of the
  Seebeck and the Peltier effect and the resulting temperature
  difference of the two reservoirs as a function of the interaction
  parameter $(k_Fa)^{-1}$.  The Peltier effect may lead to an
  additional cooling mechanism for ultracold fermionic atoms.
\end{abstract}



\maketitle

\section{Introduction}
Thermal transport is an important tool to investigate many-body
systems. There is a variety of transport coefficients describing the
heat carried by thermal currents as well as the voltages (in the case
of charged particles) or chemical potential differences (in the case
of neutral particles) induced by a thermal gradient (Seebeck
effect). The inverse effect, the build-up of a thermal gradient by a
particle current is of great practical importance (Peltier
effect). These thermoelectric effects depend in sensitive ways on the
excitation spectrum of the system close to the Fermi
surface~\cite{Staring1993,Guttman1995}. If the spectrum is
particle-hole symmetric (as it is to a good approximation in the bulk
of a metallic superconductor), the Seebeck effect vanishes. Breaking
this symmetry in superconducting tunnel junctions allows for
refrigeration~\cite{PekolaRMP2006} and/or giant thermoelectric
effects~\cite{Machon2013,Heikkilae2014,Beckmann2016}.

In recent years, transport in ultracold atomic gases has been
investigated both theoretically~\cite{Holland2007, Holland2009,
  Holland2010, Bruderer2012, Wimberger2013} 
and in a number of experiments~\cite{Brantut2012,
  Stadler2012, Brantut2013, Krinner2015, Husmann2015}. Optical
potentials were used to realize a narrow channel connecting two
macroscopic reservoirs of neutral fermionic atoms to form an atomic
analogue of a quantum mesoscopic device. Ohmic conduction in such a
setup was observed~\cite{Brantut2012} as well as conductance plateaus
at integer multiples of the conductance quantum $1/h$ for a ballistic
channel~\cite{Krinner2015}. Tuning the interaction between the atoms
by a magnetic field via a Feshbach resonance allowed to drive the
system into the superfluid regime. The resulting drop of the
resistance was observed experimentally~\cite{Stadler2012}.  Moreover,
a quantum point contact between two superfluid reservoirs was
realized~\cite{Husmann2015}. Signatures of thermoelectric effects were
observed in the normal state of these systems~\cite{Brantut2013}.
Several theoretical studies also examined mesoscopic
transport~\cite{Bruderer2012}, thermoelectric
effects~\cite{Grenier2012}, and Peltier cooling in ultracold fermionic
quantum gases~\cite{Grenier2014,Grenier2016}.

In this paper, we investigate the coupling of thermal and particle currents in a junction of two superfluids. The goal is to explore the possibility to realize dynamical heating and refrigeration phenomena around the phase transition.
To this end, we consider two reservoirs of interacting ultracold
atoms connected by a weak link that we model as a tunnel junction. 
The generalized BCS theory~\cite{Leggett2006}
provides self-consistency equations for the gap parameter and the
chemical potential as a function of the dimensionless interaction parameter
$(k_Fa)^{-1}$. We use the tunneling approach to describe quasiparticle
transport in a system with a fixed number of particles and specify the
initial particle and/or temperature imbalance of the two
reservoirs. The resulting time evolution of the system shows
a number of characteristic features: we find transitions between
superfluid and normal states as well as signatures of the Peltier and
Seebeck effects. In addition, there are peaks in the transport
current that can be related to a resonant condition in the expression
for the tunneling current.

The paper is organized as follows: In Sec.~\ref{sec:model} we
introduce a model Hamiltonian for the system consisting of two
tunnel-coupled reservoirs as well as the self-consistency equations
for the superconducting gap and the chemical potential in the
generalized BCS theory. We also give expressions for the particle and the heat
current.  In Sec.~\ref{sec:time_evo} we calculate
the time evolution of the system with a fixed total number of
particles initially prepared with an imbalance in particle number
and/or temperature. Finally, we conclude in Sec.~\ref{sec:conclusions}.

\section{Model} \label{sec:model} 
\begin{figure}[t]
\includegraphics[width=1\columnwidth]{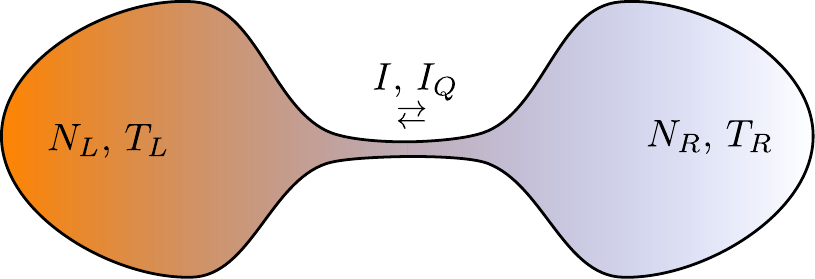}
\caption{Two reservoirs of ultracold fermions connected via a tunnel
  junction allowing particle and heat transport. Each reservoir is
  characterized by the particle number $N$ and temperature $T$.}
  \label{fig:reservoirs}
\end{figure}

Our system, depicted in Fig.~\ref{fig:reservoirs}, consists of two
reservoirs of interacting neutral fermionic atoms connected by a weak
link that is modeled by a tunnel junction. Experimentally, the junction can
be realized as a constriction in space using trapping lasers. We
denote the number of particles and temperature in the left (right)
reservoir as $N_{L(R)}$ and $T_{L(R)}$, respectively.

The Hamiltonian describing this system is assumed to be
\begin{equation}
 H = H_L + H_R +H_t\:,
\end{equation}
where $H_L$ and $H_R$ are the BCS Hamiltonians for the two reservoirs
\begin{equation}\label{eq:HL_HR}
 \begin{aligned}
H_L&=\sum_{p\sigma}\xi_p c^{\dagger}_{p\sigma}c_{p\sigma}+\frac{1}{2}\sum_{pp'\sigma}V_{pp'}
c^{\dagger}_{p\sigma}c^{\dagger}_{-p-\sigma}c_{-p'-\sigma}c_{p'\sigma}\:,
\\
H_R&=\sum_{k\sigma}\xi_k a^{\dagger}_{k\sigma}a_{k\sigma}+\frac{1}{2}\sum_{kk'\sigma}V_{kk'}
a^{\dagger}_{k\sigma}a^{\dagger}_{-k-\sigma}a_{-k'-\sigma}a_{k'\sigma}\:.
 \end{aligned}
\end{equation}
Here, $c_{p\sigma}$ and $c^\dagger_{p\sigma}$ ($a_{p\sigma}$ and
$a^\dagger_{p\sigma}$) are the annihilation (creation) operators of a fermion
with momentum $p$ and spin $\sigma$ in the left (right) reservoir,
$\xi_p=\varepsilon_p-\mu$ is the single-particle energy with respect
to the chemical potential, and $V_{pp'}$ is the (singlet) pairing
interaction. In the context of neutral fermionic atoms the spin degree
of freedom is represented by the two hyperfine states of the atom in
consideration. The tunneling Hamiltonian is
\begin{equation}\label{eq:Ht}
 H_t=\sum_{kp\sigma} \eta_{kp}a^{\dagger}_{k\sigma}c_{p\sigma}+h.c.\:,
\end{equation}
where $\eta_{kp}$ is the tunneling matrix element, which in the following
we assume to be energy independent, $|\eta_{kp}|^2=|\eta|^2$.

In the next step, we restrict ourselves to the mean-field
approximation for the Hamiltonians in Eq.~\eqref{eq:HL_HR} introducing the
mean-field parameter $\Delta_L$ for the left reservoir
\begin{equation}
\Delta_{p\sigma -\sigma}=-\sum_{p'} V_{pp'}
\left< c_{-p'-\sigma}c_{p'\sigma}\right> \approx \Delta_L
\end{equation}
and analogously for the right reservoir. 

\begin{figure}[t]
\centering
\includegraphics[width=\columnwidth]{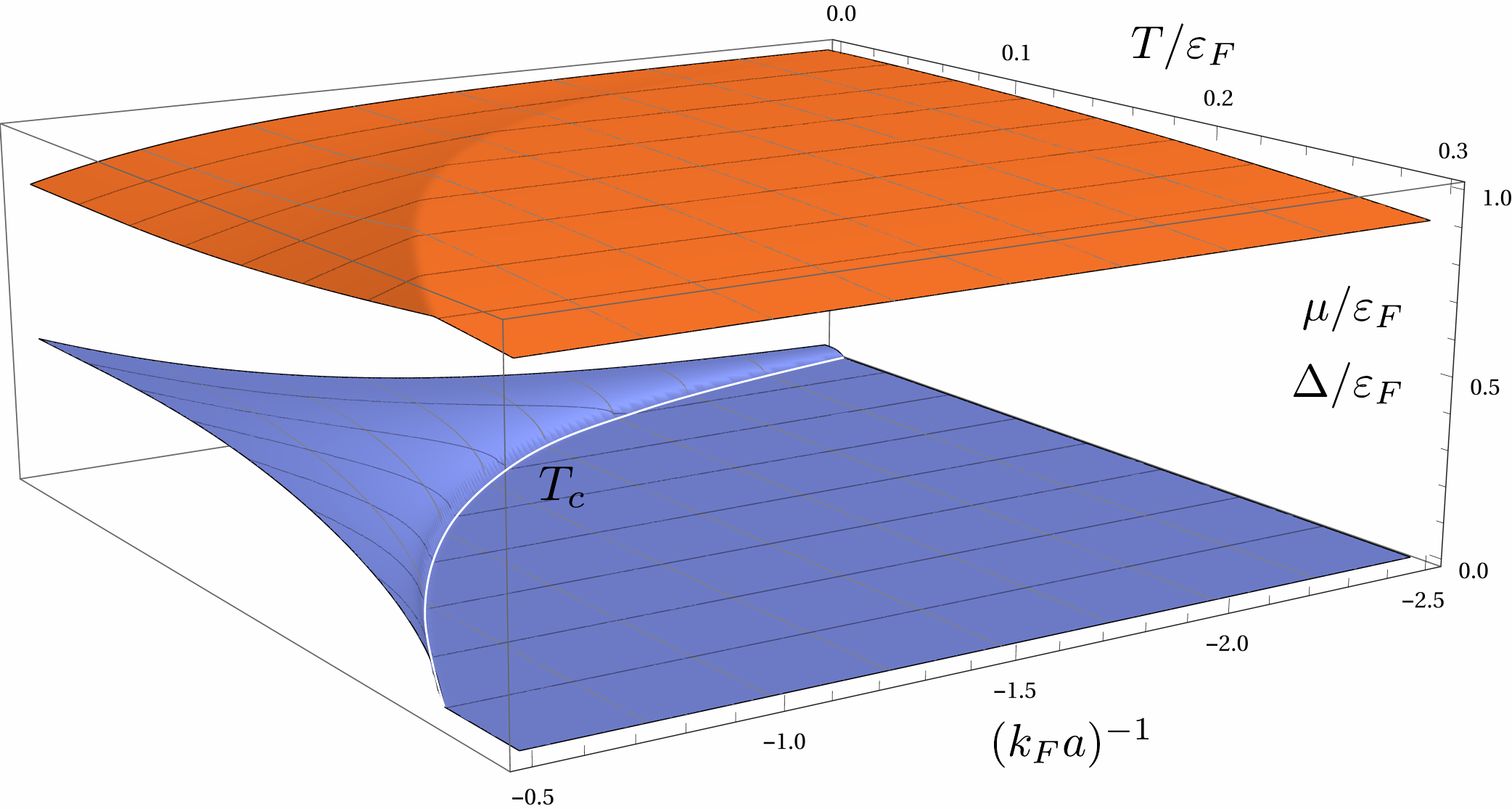}
\caption{Solution for $\Delta$ (blue) and $\mu$ (orange) following from Eqs.~\eqref{eq:gap_eq} and \eqref{eq:mu_eq} as a
  function of $(k_Fa)^{-1}$ and $T$. The mean-field critical
  temperature $T_c$ is shown as a white curve. In the BCS limit
  $(k_Fa)^{-1}\to - \infty$, the chemical potential
  $\mu/\varepsilon_F\to 1$ and $\Delta$ as well as $T_c$ approach
  zero.}
  \label{fig:selfconsistent_eqs_fig}
\end{figure}

In a dilute gas of neutral fermionic atoms it is a good approximation
to describe the interaction $V_{pp'}$ between two atoms using a single
parameter, the s-wave scattering length $a$. Consequently, the
dimensionless interaction parameter $(k_Fa)^{-1}$ can be included in
the BCS gap equation using a standard renormalization procedure (see,
e.g. Appendix 8A of Ref.~\onlinecite{Leggett2006}). The gap equation
then takes the form
\begin{equation}\label{eq:gap_eq}
\frac{\pi}{k_F a}\sqrt{\varepsilon_F} = 
\int_0^\infty \ud\varepsilon \sqrt{\varepsilon}
\left[\frac{1}{\varepsilon}-\frac{1}{E}\tanh\left(\frac{E}{2T}\right)\right],
\end{equation}
where $E=\sqrt{(\varepsilon-\mu)^2+|\Delta|^2}$ and $\varepsilon_F$ is
the Fermi energy. In Eq.~\eqref{eq:gap_eq}, there are two unknown
variables $\mu$ and $\Delta$. To solve it, the second equation is
obtained by fixing the number of particles
\begin{equation}\label{eq:mu_eq}
\frac{4}{3}\varepsilon_F^{3/2} = \int_0^\infty \ud \varepsilon
\sqrt{\varepsilon}\left[1-\frac{\varepsilon-\mu}{E}\tanh\left(\frac{E}{2T}\right)\right].
\end{equation}
For the density of states (DOS) of a 3D Fermi gas in the normal state
$\mathcal{N}^0(\varepsilon)\propto\sqrt{\varepsilon}$ (neglecting the
confining potential) which we used above, the integrals in
Eqs.~\eqref{eq:gap_eq} and \eqref{eq:mu_eq} converge and no cut-off
energy needs to be introduced.  The solution to these equations is
shown in Fig.~\ref{fig:selfconsistent_eqs_fig} as a function of
temperature $T$ and interaction parameter $(k_Fa)^{-1}$. As the
interaction parameter approaches the BCS limit,
$(k_Fa)^{-1}\ll -1$, 
the superconducting gap $\Delta$ and
critical temperature $T_c$ are proportional to $e^{-\pi/(2k_Fa)}$
and $\mu/\varepsilon_F\to 1$ at $T=0$~\cite{Leggett2006}. On the other
hand, towards unitarity, where $(k_Fa)^{-1}\to 0^-$, $\Delta$ and
$T_c$ increase and $\mu$ decreases. 

Note that this mean-field critical temperature $T_c$ is in fact the
pairing temperature below which a significant number of fermions are
bound in pairs. In the BCS limit the real critical temperature and
mean-field $T_c$ coincide, however, closer to the unitary regime, this
approximation starts to fail.

An initial state with particle number imbalance or temperature
imbalance between the left and right reservoirs will give rise to
particle and heat transport. The particle current~$I$ and energy
current~$I_\mathcal{E}$ are defined as
\begin{equation}
 \begin{aligned}
  I &= -\frac{\partial \langle\hat{N}_L\rangle}{\partial t} = i \langle[\hat{N}_L,H]\rangle \\
  I_\mathcal{E} &= -\frac{\partial \langle H_L\rangle}{\partial t} = i \langle[ H_L,H]\rangle\:,\\
 \end{aligned}
\end{equation}
where the angular brackets represent the thermodynamic average in the
grandcanonical ensemble and
$\hat{N}_L=\sum_{p\sigma}c^\dagger_{p\sigma}c_{p\sigma}$ is the
fermion number operator in the left reservoir. All the operators are
in the Heisenberg picture.

If we restrict ourselves to quasiparticle transport (ignoring Cooper
pairs and interference terms between Cooper pairs and quasiparticles),
the expressions for the particle and heat current in the tunneling
limit read
\begin{align}
\label{eq:particle_current}
 I &= I_{L\to R}-I_{R\to L}\\
 &= \frac{2\pi|\eta|^2}{\hbar} \mathcal{V}_L\mathcal{V}_R
   \int_{-\infty}^\infty \ud E \mathcal{N}_L(E)\mathcal{N}_R(E)\left[
   f_L(E)-f_R(E) \right] \nonumber
\end{align}
and
\begin{align}\label{eq:heat_current}
 I_Q &= I_{Q,L\to R} - I_{Q,R\to L} \nonumber\\ 
 &=\frac{2\pi|\eta|^2}{\hbar} \mathcal{V}_L\mathcal{V}_R \int_{-\infty}^\infty \ud E \mathcal{N}_L(E)\mathcal{N}_R(E)\nonumber\\
 &\times \left[(E-\mu_L)f_L(E)(1-f_R(E))\right.\nonumber\\
& \left.-(E-\mu_R)f_R(E)(1-f_L(E))\right]\:.
\end{align}
Here, $\mathcal{V}_{L(R)}$ is the volume 
and $f_{L(R)}(E)$ the Fermi function
describing the left (right) reservoir.
The superconducting density of states
\begin{equation*}
\mathcal{N}_{L(R)}(E)=\text{Re}\,\{\mathcal{N}^0_{L(R)}(\varepsilon)\} \text{Re}\,\{ \frac{|E-\mu_{L(R)}|}{\sqrt{(E-\mu_{L(R)})^2-\Delta^2_{L(R)}}} \}
\end{equation*}
contains the energy-dependent density of states $\mathcal{N}^0_{L(R)}$
of a normal 3-dimensional Fermi gas 
that can be expressed as 
\begin{align*}
&\mathcal{N}^0_{L(R)}(\varepsilon)=\frac{1}{2\pi^2}(\frac{2m}{\hbar^2})^{3/2}\sqrt{\varepsilon}
  = \frac{1}{2\pi^2}(\frac{2m}{\hbar^2})^{3/2}\\
&\times\sqrt{\mu_{L(R)}+\text{sign}(E-\mu_{L(R)})\text{Re}\sqrt{(E-\mu_{L(R)})^2-\Delta^2_{L(R)}}}\:.
\end{align*}

\section{Time evolution of the system} 
\label{sec:time_evo} 
For finite reservoirs, which is the case we are studying here, a
non-equilibrium initial state (like a temperature or particle number
imbalance between the left and right reservoir) will induce
time-dependent transport~\cite{Bruderer2012, Grenier2012,
  Grenier2014}. To model this phenomenon we consider the balance
equations for the particle number $N_{L(R)}$ and energy
$\mathcal{E}_{L(R)}$ in each reservoir that lead to
\begin{equation}\label{eq:time_evo}
 \begin{aligned}
  \frac{\partial N_{L(R)}}{\partial t} &= \mp I \\
   \frac{\partial T_{L(R)}}{\partial t} &=\mp \frac{1}{C_{\mathcal{V}_{L(R)}}}(I_Q+\mu_L I_{L\to R}-\mu_R I_{R\to L})\:.
 \end{aligned}
\end{equation}
Here, we used the relation between the energy
of the left (right) reservoir and temperature change of the system at
constant volume $C_{\mathcal{V}}=\partial \mathcal{E}/\partial T$. The
heat capacity in the BCS theory is given by
\begin{equation}
 \begin{aligned}\label{eq:spec_heat}
  C_{\mathcal{V}}(T) &=  \frac{2}{T}\int_{-\infty}^\infty \ud E \mathcal{N}(E) \left(-\frac{\partial f(E)}{\partial E}\right) \\
 &\times\left(E^2-\frac{T}{2}\frac{\partial \Delta^2}{\partial T}+T\text{sign($E$)}\sqrt{E^2-\Delta^2}\frac{\partial\mu}{\partial T}\right)\:.
 \end{aligned}
\end{equation}
In writing Eqs.~(\ref{eq:time_evo}) and (\ref{eq:spec_heat}), we have neglected number and
energy fluctuations in the reservoirs which were shown to be small in
the regime considered here~\cite{Schroll2007}.

To calculate the time evolution of the system, we proceed as follows:
starting with $N_{L(R)}(t)=N\pm\delta N /2$ and
$T_{L(R)}(t)=T\pm\delta T /2$ at time $t$, we calculate the
corresponding values of $\mu_{L(R)}(t)$ and $\Delta_{L(R)}(t)$ using
Eqs.~\eqref{eq:gap_eq} and \eqref{eq:mu_eq}. Then, using the
discretized form of Eq.~\eqref{eq:time_evo}, we obtain
$N_{L(R)}(t+\delta t)$ and $T_{L(R)}(t+\delta t)$ at time
$t+\delta t$, and the procedure is iterated. The time evolution is
hence uniquely determined by setting initial values of $N^0_{L(R)}$,
$T^0_{L(R)}$ and $(k^0_{F,L}a)^{-1}$,
where quantities with superscript $0$ denote the values 
at time $t=0$.  The interaction parameter on the right side follows
from $(k^0_{F,L}a)^{-1}$ and $N^0_{R}$. 
Note that in linear response in $\delta N$ and
$\delta T$, assuming $\Delta_L=\Delta_R=0$ and $C_\mathcal{V}=$
constant, Eqs.~\eqref{eq:time_evo} can be solved analytically using
simple exponential functions\cite{Grenier2012}. For example, an initial
particle number imbalance will decay exponentially with time.

Typically, starting with an initial particle number (temperature)
imbalance $\delta N_0$ ($\delta T_0$) will lead to a time-dependent
temperature (particle number) imbalance due to the coupling between
particle and heat transport. As a consequence, the chemical potential
imbalance $\delta\mu=\mu_L-\mu_R$ and $\delta\Delta=\Delta_L-\Delta_R$
will also depend on time.  Eventually, as $t\to\infty$, the system
reaches an equilibrium state.

\begin{figure}[t]
\centering
\includegraphics[width=\columnwidth]{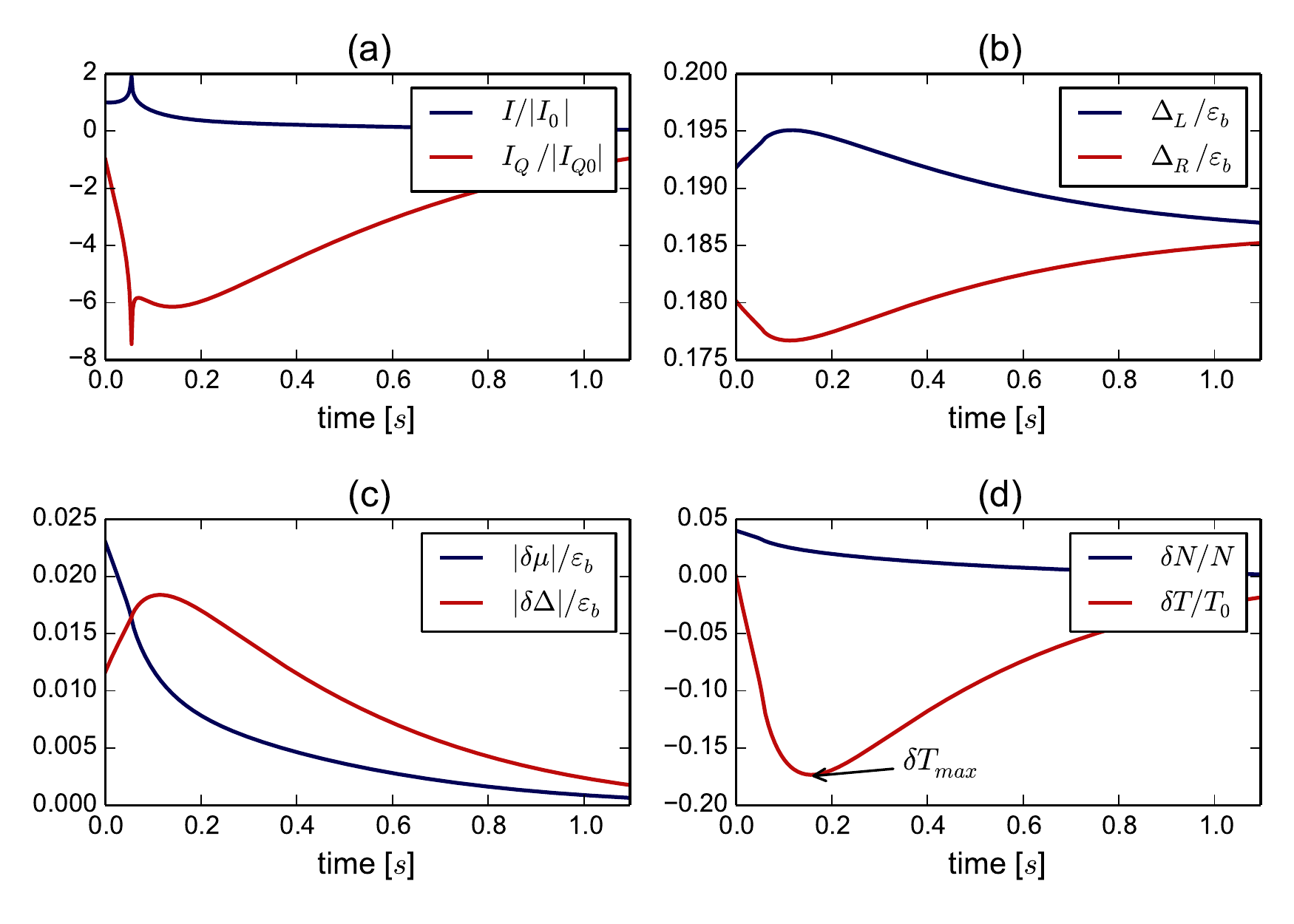}
\caption{Time evolution of various quantities: (a) particle and heat
  current. (b) superconducting gap in the left and right
  reservoir. (c) chemical potential difference and difference between
  gaps in the left and right reservoir. (d) particle number difference
  and temperature difference. The sharp peak in the currents occurs
  for the time $t$ at which $|\delta\mu|=|\Delta_L-\Delta_R|$, {\it
    i.e.}, when thermally excited quasiparticles are allowed to tunnel
  between the peaks in the DOS of the two reservoirs. The initial
  conditions chosen are $N=2\times 10^4$, $\delta N_0/N=0.04$,
  $T^0_L=T^0_R=T_0=0.07\varepsilon_b$, and
  $(k^0_{F,L}a)^{-1}=-1$.}
  \label{fig:time_evo1}
\end{figure}

\begin{figure}[t]
\centering
\includegraphics[width=\columnwidth]{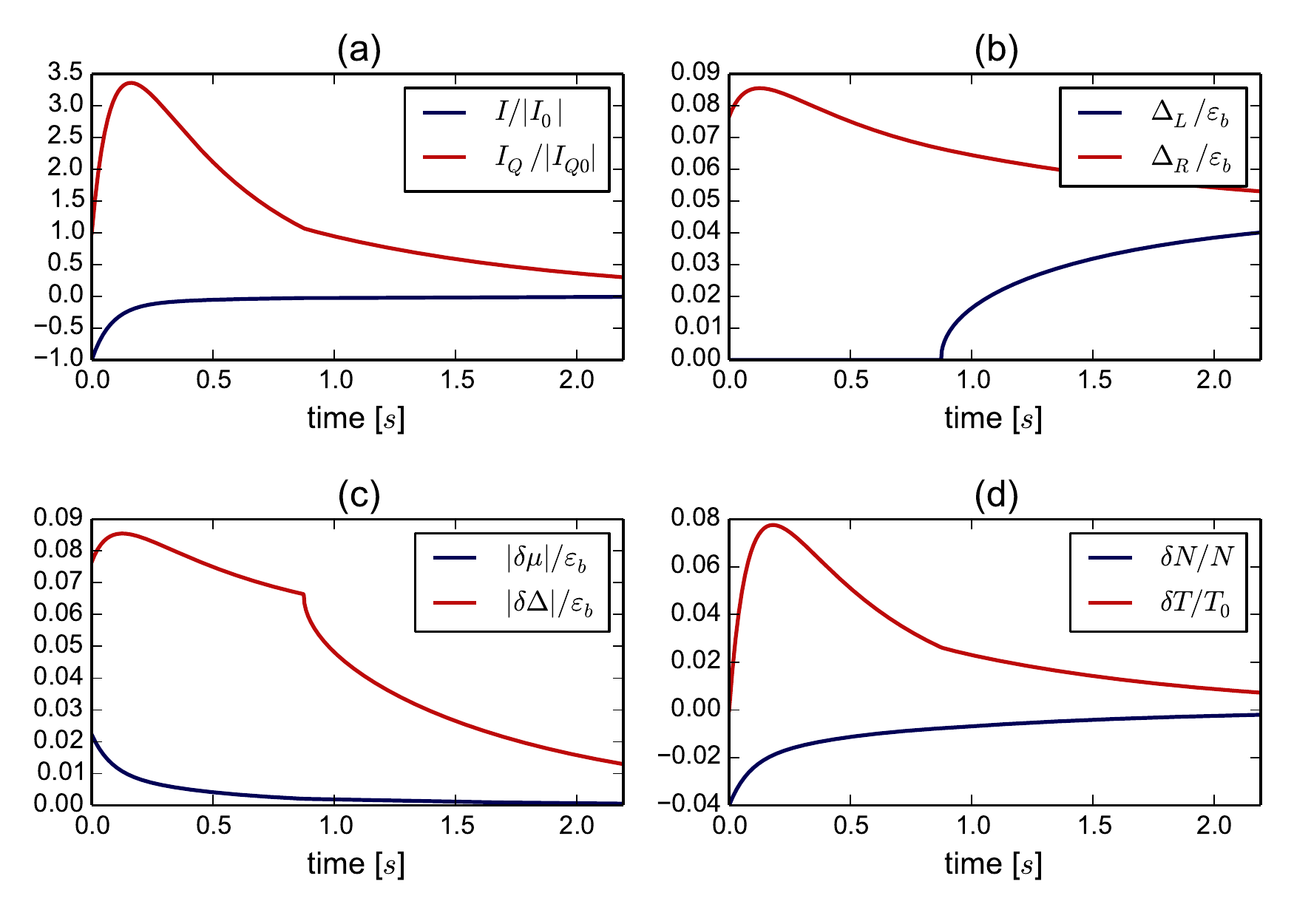}
\caption{Time evolution of the same quantities as in
  Fig.~\ref{fig:time_evo1}. A negative initial particle number
  imbalance and an initial temperature between the transition
  temperatures of the two reservoirs leads to a transition of the left
  reservoir from an initially normal to a superfluid state at
  intermediate times. The initial conditions are $N=2\times 10^4$,
  $\delta N_0/N= -0.04$, $T^0_L=T^0_R=T_0=0.1248\,\varepsilon_b$, and
  $(k^0_{F,L}a)^{-1}=-1$.  }
  \label{fig:time_evo2}
\end{figure}

\begin{figure}[t]
\centering
\includegraphics[width=\columnwidth]{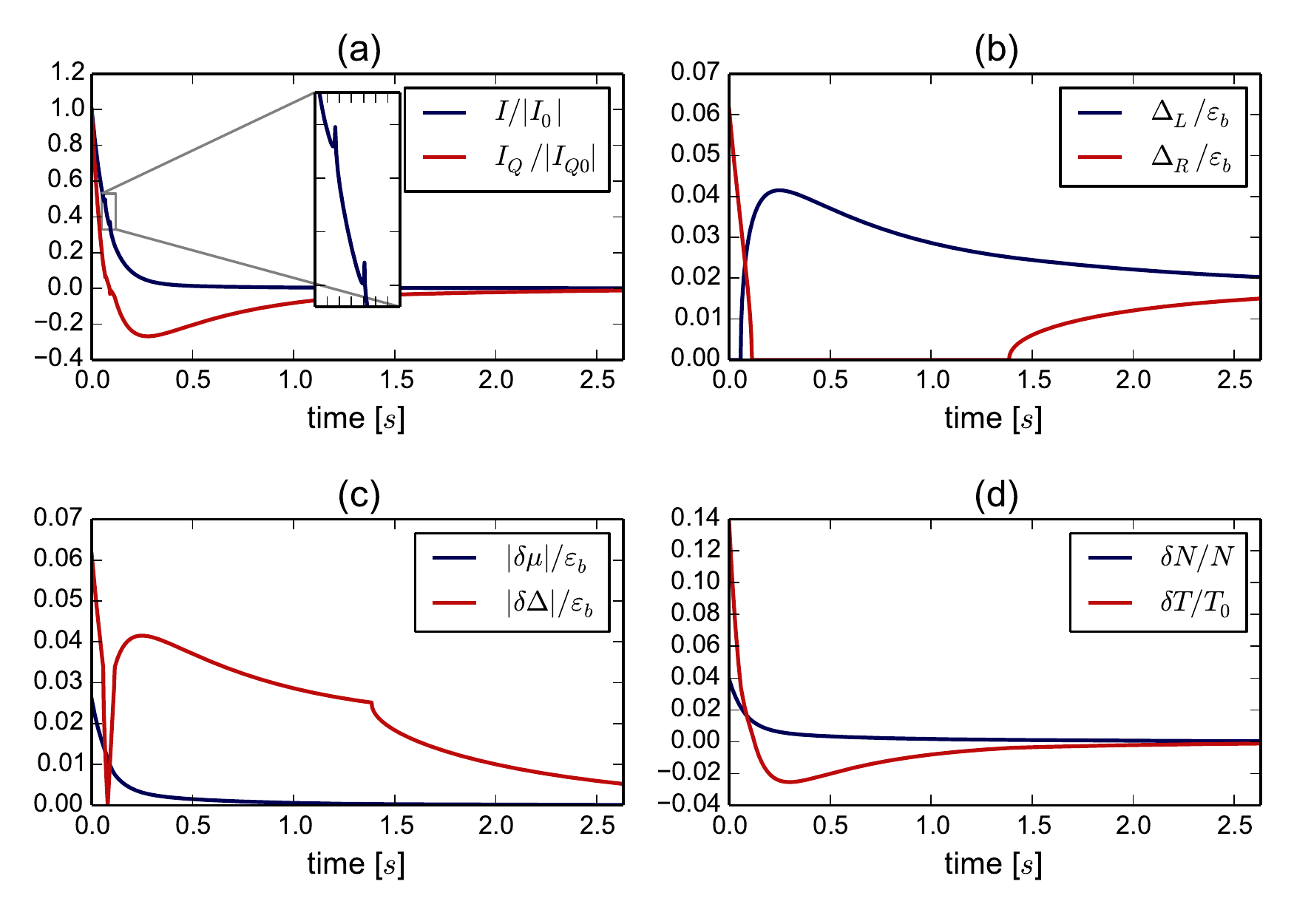}
\caption{Time evolution of the same quantities as in
  Fig.~\ref{fig:time_evo1}. The system exhibits several
  transitions. The peaks in the particle and heat current are present
  for the same reason as in Fig.~\ref{fig:time_evo1}. In this case the
  condition $|\delta\mu|=\left| \Delta_L-\Delta_R \right|$ is
  satisfied twice during the time evolution. The initial conditions
  are $N=2\times 10^4$, $\delta N_0/N= 0.04$,
  $T^0_L=0.132\,\varepsilon_b$, $T^0_R=0.115\,\varepsilon_b$,
  $T_0=(T^0_L+T^0_R)/2$, and $(k^0_{F,L}a)^{-1}=-1$.}
  \label{fig:time_evo3}
\end{figure}

In the following we show and discuss three examples of such a time
evolution displaying various quantities characterizing the system as a
function of time. The time scale in
Figs.~\ref{fig:time_evo1}--\ref{fig:time_evo3} is fixed as follows:
time can be expressed in units of $\varepsilon_b\hbar/|\eta|^2$, where
$\varepsilon_b=\hbar^2/(2ma^2)$ and $|\eta|^2=|\eta_{kp}|^2$ is the
modulus squared of the tunneling matrix element introduced after
Eq.~\eqref{eq:Ht}. As mentioned earlier, the time evolution of a
system in the normal state within linear response corresponds to an
exponential decay of the initial particle number imbalance. To get an
order-of-magnitude estimate for the absolute time scale in seconds, 
we compare our results for the dimensionless linear response
coefficient $1/\tilde{\tau}$ in $\tilde{I} = \delta N/\tilde{\tau}$,
where the tilde denotes dimensionless quantities, with the
experimental value $1/\tau_0=2.9\,\mathrm{s}^{-1}$ taken from
Ref.~\onlinecite{Brantut2012}. This leads to relation
\begin{equation*}
\frac{\varepsilon_b \hbar}{|\eta|^2} =  \tau_0/\tilde{\tau}.
\end{equation*}
The time scale $\tau_0$ represents a characteristic particle
transport time scale and is analogous to the $RC$-time of a capacitor
circuit.

Figure~\ref{fig:time_evo1} demonstrates a case in which a sharp peak
in the current as a function of time appears. This can be understood in
the semiconductor picture of the tunneling process: the BCS DOS at the
edges of the gap, $E=\pm\Delta$, in both reservoirs is divergent,
provided that both reservoirs are in the superfluid regime. Hence, if
the condition $|\delta\mu(t)|=|\Delta_L(t)-\Delta_R(t)|$ is satisfied,
electrons from a peak in the DOS of one reservoir are allowed to
tunnel into the peak in the DOS of the other reservoir. This condition
creates a logarithmic singularity in the integrals in
Eqs.~\eqref{eq:particle_current},~\eqref{eq:heat_current} (in the
absence of gap anisotropy and level broadening)~\cite{Tinkham2004}.
Moreover, a time-dependent temperature imbalance $\delta T(t)$
develops that exhibits a non-monotonic behavior and reaches its
maximum value $\delta T_\text{max}$ at a certain time, see
Fig.~\ref{fig:time_evo1}(d).  The build-up of this temperature
imbalance is a signature of the Peltier effect. For
the case shown in Fig.~\ref{fig:time_evo1} the initial conditions are
chosen such that both reservoirs are in the superfluid regime
throughout the time evolution: $N=2\times 10^4$, $\delta N_0/N=0.04$,
$T^0_L=T^0_R=T_0=0.07\,\varepsilon_b$, $(k^0_{F,L}a)^{-1}=-1$. 
The corresponding initial values of $T_c^0$ are 
$T_{c,L}^0=0.125\,\varepsilon_b$ and $T_{c,R}^0=0.119\,\varepsilon_b$.

In Fig.~\ref{fig:time_evo2} we choose a negative initial particle
number imbalance $\delta N_0/N= -0.04$ (while keeping
$(k^0_{F,L}a)^{-1}=-1$) and an initial temperature
$T^0_L=T^0_R=T_0=0.1248\,\varepsilon_b$ that lies between the
initial transition temperatures of the two reservoirs.
Since $T_{c,L}^0=0.119\,\varepsilon_b$ and $T_{c,R}^0=0.125\,\varepsilon_b$
in this case, the left reservoir is initially normal and the right one
superfluid. During the time evolution, the left reservoir
undergoes a transition to a superfluid state as shown in 
Fig.~\ref{fig:time_evo2}(b). Interestingly, this is not caused by
lowering the temperature in the left reservoir. On the contrary, the
temperature in the left reservoir actually temporarily rises. But the
particle number (and hence the density) in the left reservoir rises
which causes the transition from $\Delta_L=0$ to $\Delta_L\neq 0$. 
As before, the calculation was done for $N=2\times 10^4$.

Figure~\ref{fig:time_evo3} shows a more complex time evolution. The
peaks in the current as a function of time appear for the same reason
as in Fig.~\ref{fig:time_evo1}(a), but now the condition
$|\delta\mu|=|\Delta_L(t)-\Delta_R(t)|$ is satisfied twice during the
time-evolution, see Fig.~\ref{fig:time_evo3}(c). The system also
undergoes several superfluid transitions similar to
Fig.~\ref{fig:time_evo2}(b). Finally, when the system equilibrates for
$t\to\infty$, both reservoirs end up in the superfluid state. The
initial conditions were chosen as $N=2\times 10^4$,
$\delta N_0/N= 0.04$, $T^0_L=0.132\,\varepsilon_b$,
$T^0_R=0.115\,\varepsilon_b$, $T_0=(T^0_L+T^0_R)/2$, and $(k^0_{F,L}a)^{-1}=-1$.

\begin{figure}[t]
\centering
\includegraphics[width=0.9\columnwidth]{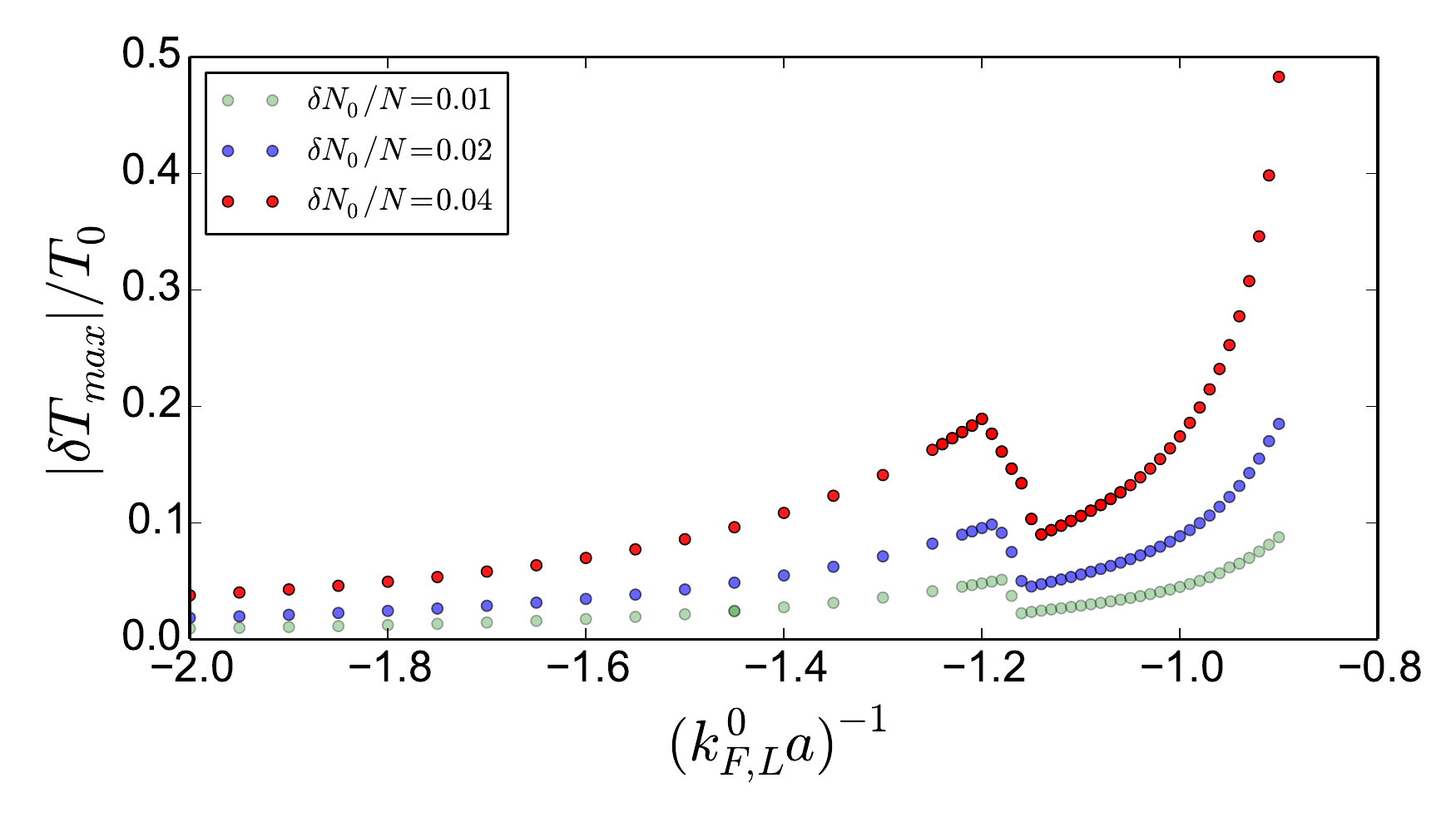}\\
\includegraphics[width=0.9\columnwidth]{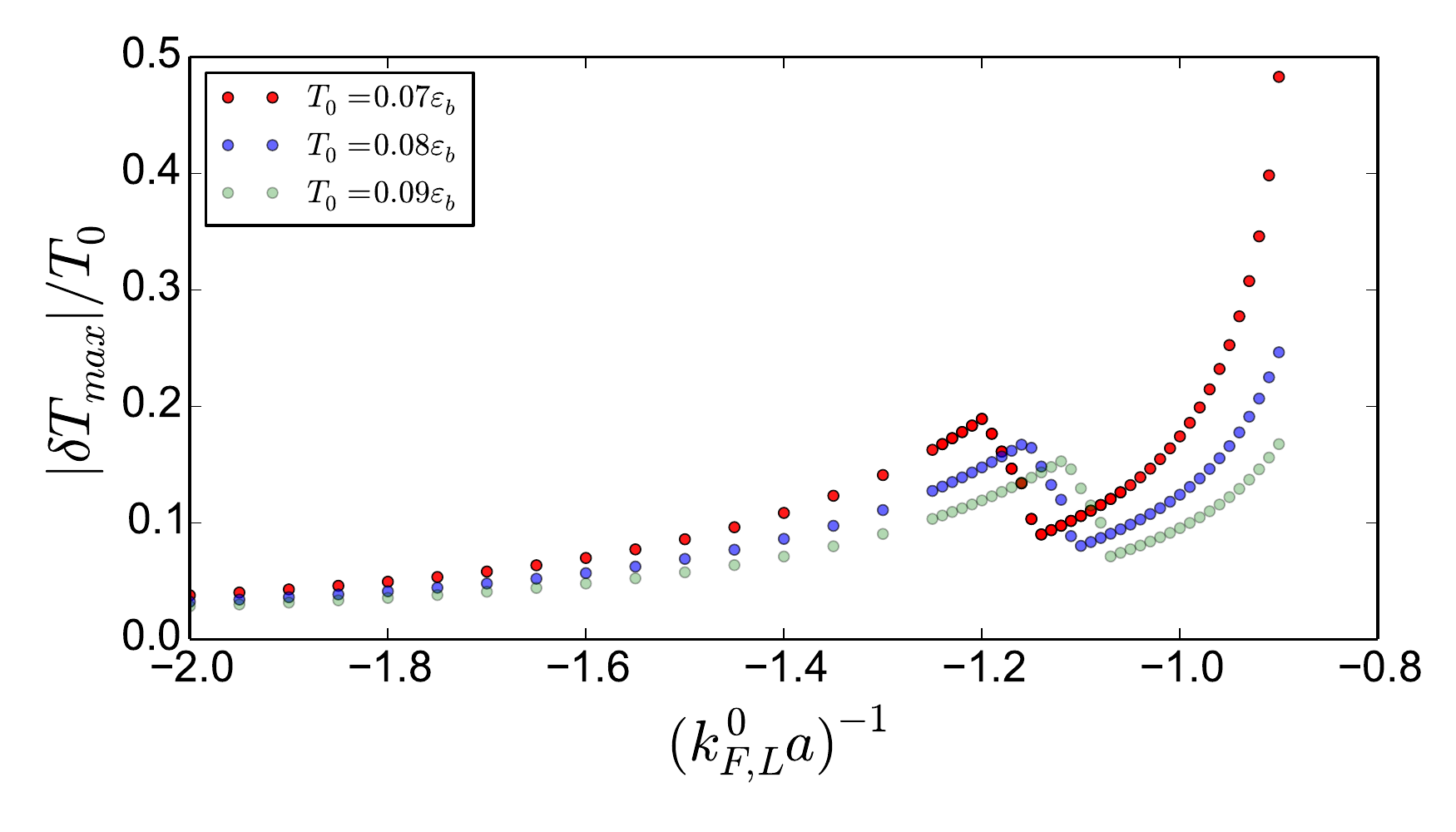}
\caption{Maximal induced temperature imbalance $|\delta T_\text{max}|$
  as a function of $(k_{F,L}^0a)^{-1}$ for different values of the
  initial particle number imbalance $\delta N_0$ and initial
  temperature $T^0_L=T^0_R=T_0$.  Upper panel: $N=2\times 10^4$,
  $T_0=0.07\,\varepsilon_b$, and three different values of
  $\delta N_0/N$. Lower panel: $N=2\times 10^4$, $\delta N_0/N=0.04$,
  and three different values of $T_0$.  The Peltier effect gets more
  significant approaching the unitary point.  }
\label{fig:tmax_v_kFa}
\end{figure}

As mentioned earlier, the induced temperature imbalance $\delta T$ due
to an initial particle number imbalance $\delta N_0$ is a signature of
the Peltier effect.  It shows a non-monotonous behavior as a function
of time with a maximum $\delta T_\text{max}$ at intermediate times,
see Figs.~\ref{fig:time_evo1}(d) and \ref{fig:time_evo2}(d).  In
Fig.~\ref{fig:tmax_v_kFa} we show $|\delta T_\text{max}|$ as a
function of $(k^0_{F,L}a)^{-1}$ for different values of the initial
particle number imbalance $\delta N_0$ and initial temperature
$T^0_L=T^0_R=T_0$. Each of the functions is divided into two sections
monotonically increasing with increasing $(k^0_{F,L}a)^{-1}$. The left
section represents data from a system which is in the normal state,
$\Delta_{L(R)}(t) = 0$, during the whole time evolution, whereas for
the right section $\Delta_{L(R)}(t) \neq 0$, as in
Fig.~\ref{fig:time_evo1}. Between the two sections, there is a
``transient'' regime, where superfluid transitions occur, similar to
the ones in Figs.~\ref{fig:time_evo2} and \ref{fig:time_evo3}. The
increase of $|\delta T_\text{max}|$ towards unitarity cannot be
explained by particle-hole asymmetry alone but is due to a delicate
interplay of the various factors in the integrands of
Eqs.~\eqref{eq:particle_current} and \eqref{eq:heat_current}.

\section{Conclusion} 
\label{sec:conclusions}
To summarize, we have investigated particle and heat transport on the
BCS side of the BCS-BEC crossover in a two-terminal setup with two
reservoirs of interacting ultracold atoms. We have shown that a system
initially out of equilibrium will show particle and/or thermal
currents whose existence leads to characteristic time-dependent
signatures, such as transitions between normal and superconducting
states and resonant features in the currents as a function of time.
An initial temperature imbalance can lead to a difference in chemical
potentials at intermediate times. This is a signature of the Seebeck
effect.  Conversely, an initial particle number imbalance for two
reservoirs at equal temperatures can lead to the build-up of a
temperature difference at intermediate times, which is a signature of
the Peltier effect.  The maximal induced temperature imbalance
increases if $(k_Fa)^{-1}$ moves closer to the unitarity limit.

In conclusion, our paper points out a variety of dynamical features
visible in the equilibration process that can be used to pin-point the
parameters of the system. An experimental confirmation of the Peltier
effect discussed here is important since an additional cooling
mechanism for ultracold fermionic atoms will be a valuable resource.
Furthermore, transport experiments in systems of ultracold atoms provide a
fascinating laboratory in which the combination of particle and
thermal currents can be explored in a regime that is not accessible to
experiments with metallic superconductors.

\acknowledgments
TS and CB acknowledge financial support by the Swiss SNF and the
NCCR Quantum Science and Technology. WB was financially supported by
the DFG through SFB 767.


\begin{thebibliography}{99}

\bibitem{Staring1993}
A.A.M. Staring, L.W. Molenkamp, B.W. Alphenaar, H. van Houten,
O.J.A. Buyk, M.A.A. Mabesoone, 	C.W.J. Beenakker, and C.T. Foxon, 
Europhys. Lett. \textbf{22}, 57 (1993).

\bibitem{Guttman1995} G.D. Guttman, E. Ben-Jacob, and D.J. Bergman, 
Phys. Rev. B \textbf{51}, 17758 (1995).

\bibitem{PekolaRMP2006} F. Giazotto, T.T. Heikkil\"a, A. Luukanen, A.M. Savin, and J.P. Pekola,
Rev. Mod. Phys. {\bf 78}, 217 (2006).

\bibitem{Machon2013} P. Machon, M. Eschrig, and W. Belzig,
Phys. Rev. Lett. {\bf 110}, 047002 (2013).

\bibitem{Heikkilae2014} A. Ozaeta, P. Virtanen, F.S. Bergeret, and
  T.T. Heikkil\"a, 
Phys. Rev. Lett. {\bf 112}, 057001 (2014).

\bibitem{Beckmann2016} S. Kolenda, M.J. Wolf, and D. Beckmann,
Phys. Rev. Lett. {\bf 116}, 097001 (2016).

\bibitem{Holland2007} B.T. Seaman, M. Kr\"amer, D.Z. Anderson, and M.J. Holland,
Phys. Rev. A {\bf 75}, 023615 (2007).

\bibitem{Holland2009} R.A. Pepino, J. Cooper, D.Z. Anderson, and M.J. Holland,
Phys. Rev. Lett. {\bf 103}, 140405 (2009).

\bibitem{Holland2010} R.A. Pepino, J. Cooper, D. Meiser,
  D.Z. Anderson, and M.J. Holland,
Phys. Rev. A {\bf 82}, 013640 (2010).

\bibitem{Wimberger2013} A. Ivanov, G. Kordas, A. Komnik, and
  S. Wimberger,
Eur. Phys. J. B {\bf 86}, 345 (2013).

\bibitem{Brantut2012}
J.-P. Brantut, J. Meineke, D. Stadler, S. Krinner, and T. Esslinger,
Science {\bf 337}, 1069 (2012).

\bibitem{Stadler2012}
D. Stadler, S. Krinner, J. Meineke, J.-P. Brantut, and T. Esslinger,
Nature {\bf 491}, 736 (2012).

\bibitem{Brantut2013}
J.-P. Brantut, C. Grenier, J. Meineke, D. Stadler, S. Krinner,
C. Kollath, T. Esslinger, and A. Georges,
Science {\bf 342}, 713 (2013).

\bibitem{Krinner2015}
S. Krinner, D. Stadler, D. Husmann, J.-P. Brantut, and T. Esslinger, 
Nature {\bf 517}, 64 (2015).

\bibitem{Husmann2015}
D. Husmann, S. Uchino, S. Krinner, M. Lebrat, T. Giamarchi,
T. Esslinger, and J.-P. Brantut,  
Science {\bf 350}, 1498 (2015).

\bibitem{Bruderer2012} M. Bruderer and W. Belzig, 
Phys. Rev. A {\bf 85}, 013623 (2012).

\bibitem{Grenier2012} C. Grenier, C. Kollath, and A. Georges, 
{\it Probing thermoelectric transport with cold atoms},
arXiv:1209.3942

\bibitem{Grenier2014} C. Grenier, A. Georges, and C. Kollath,
Phys. Rev. Lett. {\bf 113}, 200601 (2014).

\bibitem{Grenier2016} C. Grenier, C. Kollath, and A. Georges, 
{\it Thermoelectric transport and Peltier cooling of cold atomic gases},
arXiv:1607.03641

\bibitem{Leggett2006} A.J. Leggett, {\it Quantum Liquids} 
(Oxford University Press, Oxford, 2006).

\bibitem{Schroll2007} W. Belzig, C. Schroll, and C. Bruder,
Phys. Rev. A {\bf 75}, 063611 (2007).

\bibitem{Tinkham2004} M. Tinkham, {\it Introduction to Superconductivity} 
(Dover Publications, New York, 2004).

\end{thebibliography}
\end{document}